\documentclass[11pt]{article}
\usepackage[a4paper, margin=20mm]{geometry}

\usepackage[english]{babel}
\usepackage[utf8]{inputenc}

\usepackage{titlesec}
\usepackage{titling}
\usepackage{amsmath}
\usepackage{amsfonts}
\usepackage{amssymb}
\usepackage{hyperref}
\usepackage{cleveref}
\usepackage{csquotes}
\usepackage[labelfont=bf]{caption}
\usepackage{here}
\usepackage{subcaption}
\usepackage{wrapfig}
\usepackage{xurl}
\usepackage{pdflscape}
\usepackage{afterpage}
\usepackage{multicol}
\usepackage{bm}
\usepackage{color}

\usepackage[semicolon]{natbib}

\Crefname{figure}{Fig.}{Figs.}
\addto\captionsenglish{}
\usepackage{endnotes}

\usepackage{xcolor}
\hypersetup{
	colorlinks = true,
	urlcolor = black,
	linkcolor = blue,
	citecolor = blue,
}

\usepackage{graphicx}
\graphicspath{{./img/}}

\title{\textbf{Did Mary Shelley Write \textit{Frankenstein}? A Stylometric Analysis}}
\date{\vspace{-10ex}}

\author{%
  Lee Suddaby \qquad Gordon J. Ross\\[-0.2em]
  \small School of Mathematics, University of Edinburgh
}

\begin{document}

\maketitle

\vspace{7mm}

\subsection*{\centering Abstract}
\begin{displayquote}
	The novel \emph{Frankenstein} was published anonymously in 1818, and was first credited to Mary Shelley in a French translation of 1821. Since its publication, several claims - both contemporaneous and recent - have been made suggesting that \emph{Frankenstein} was actually written by Mary's husband, Percy Bysshe Shelley. We review the background of this controversy and then apply modern techniques from computational stylometry to determine who the true author is. Based on our analysis, we find extremely substantial evidence that Mary Shelley is indeed the true author of \emph{Frankenstein}, and that is it very improbable that Percy Bysshe Shelley played a heavy role in composing the text. While our finding confirms mainstream scholarly opinion regarding \emph{Frankenstein}, our analysis is the first application of stylometric techniques to this question and  provides strong objective grounds for favouring Shelley by freeing the question from some of the politics which have traditionally accompanied it.

\end{displayquote}
\hrulefill

\begin{multicols}{2}
	
\section{Background}

The novel ``\textit{Frankenstein, or The Modern Prometheus}'' (henceforth referred to as simply \textit{Frankenstein}) was published anonymously on January 1\textsuperscript{st}, 1818, and tells the story of the scientist Victor Frankenstein, who creates and animates his monster, and suffers terrible consequences as a result \citep{Shelley1818}. A French edition of the novel was published in 1821, and was the first to credit Mary Shelley as the author. Subsequent second and third English editions were published in 1823 and 1831 \citep{FrankEditions}. The 1831 edition was significantly revised from the original 1818 edition and is the edition that is most widely read today \citep[pp. li-lv]{Shelley2019}.

The accepted story of \emph{Frankenstein}'s authorship comes from the introduction to the 1831 edition \citep{Shelley1831,Shelley2019} and is as follows: In June 1816, a ghost-story writing contest took place at Lord Byron's residence by Lake Geneva in Switzerland. The contest consisted of Mary Shelley, Percy Bysshe Shelley, Lord Byron, and his physician, John William Polidori. After several days of being unable to think of a story, Mary states that the story of \textit{Frankenstein} came to her in a ``waking dream'', and from this initial story, she was encouraged by Percy Shelley to develop it into a fully-fledged novel, which she eventually published anonymously.

Publishing anonymously (or under a pen name) was typical for female writers at the time \citep{Irvine2005}. This is the process followed in the publication of novels by Jane Austen \citep{Irvine2005} and Frances Burney \citep{Burney2012}. So while the initial anonymous publication of \emph{Frankenstein} is perfectly consistent with Mary Shelley's authorship given the period, it has allowed for speculation that \emph{Frankenstein} may in fact have been written by a different author.

In March 1818, Sir Walter Scott  claimed that \emph{Frankenstein} may have been written by Mary's husband, Percy Bysshe Shelley, by \citep{Scott1818}. Following the publication of Mary Shelley's second novel, \textit{Valperga}, an anonymous 1824 review of the novel in \emph{Knight's Quarterly} noted substantial differences between the quality of \emph{Frankenstein} and its successor, claiming ``there is not the slightest trace of the same hand'', and also suggesting that Percy Shelley wrote \emph{Frankenstein} and that Mary was only responsible for \emph{Valperga} \citep{Knight1824}. 

These suppositions were all denied by Percy Shelley --- from the start, he claimed to only have been the editor of \emph{Frankenstein} rather than its author. However, it later transpired that he had written the preface of the novel as if he were Mary Shelley, as well as the poem \emph{Mutability}, which was included in the novel uncredited, which was likely a factor in leading people to believe he was the author of the novel as a whole.

More recently, John Lauritsen's 2007 book \emph{The Man Who Wrote Frankenstein} promotes the theory of Percy Shelley's authorship \citep{Lauritsen2007}, as do other writings from the same author \citep{Lauritsen2018Web,Lauritsen2018}. Others have made similar claims \citep{deHart2013,JonesUnpublished,Zimmerman1998}. Although these claims were not - and still are not - taken seriously by mainstream scholars of Shelley's work and Romanticism, they did receive substantial media attention, with Lauritsen's book receiving the largest share of this attention.

Similar to the review in \emph{Knight's Quarterly Review} published two centuries earlier, Lauritsen believes there is a substantial difference in quality between \emph{Frankenstein} and other works of Mary Shelley - particularly \emph{Valperga} and \emph{The Last Man} \citep{Lauritsen2018}.

However, textual evidence and analysis is not the only avenue that may be - and indeed has been - investigated in determining the true author of \emph{Frankenstein}. One can also consider historical evidence. Indeed, it is the analysis of handwriting in the original manuscript which has fuelled speculation over Percy Shelley's contributions \citep{Wu2015}.

In 1996, Charles E. Robinson produced transcriptions of the original manuscript, allowing the novel to be seen in its earliest draft, and for the authorship of each part of the text to be identified down to the level of individual words. Contrary to previous analysis in \citet{Murray1978}, which concluded that Percy Shelley contributed a `thousand or so' words to the rough draft, from the introduction to his 2008 book, \emph{The Original Frankenstein}, Robinson concludes that:
\begin{displayquote}
	\vspace{1pt}\vspace{-1pt}[Percy] contributed at least 4,000 to 5,000 words to this 72,000 word novel. Despite the number of Percy’s words, the novel was conceived and mainly written by Mary Shelley, as attested not only by others in their circle (e.g. Byron, Godwin, Claire and Charles Clairmont, Leigh Hunt) but by the nature of the manuscript evidence in the surviving pages of the Draft. \citep{Robinson2008}
\end{displayquote}

This so-called `handwriting argument' assumes that everything in the manuscript written in Mary Shelley's hand was originally composed by her. However, while there is no real evidence to suggest that the words in Percy Shelley's handwriting are likely to have been composed by someone other than himself,\endnote{i.e. while Mary is known to have done copy work for her husband, Lord Byron, and Thomas Love Peacock, there is no evidence of Percy Shelley doing this sort of work for anyone other than himself.} there are manuscripts in existence in Mary Shelley's handwriting of works composed by her husband, for example, his 1820 lyrical drama \emph{Prometheus Unbound} \citep{PromUnbound1991}, and the poems \emph{The Mask of Anarchy}, \emph{The Witch of Atlas}, \emph{The Cenci}, and \emph{The Sensitive Plant} \citep{ReimanPowers1977}. Therefore, a possible theory is that the same arrangement was true in the composition of \emph{Frankenstein}. However, the fact that Mary is known to have transcribed manuscripts for her husband is by no means indicative of the same arrangement having been used in the composition of \emph{Frankenstein} - a view shared by mainstream scholars.

At the very least, analysis of handwriting provides us with a lower bound to the extent of Percy Shelley's contribution, but the question as to whether he could have been the intellectual force behind the novel as a whole remains somewhat open. Was his contribution to the novel merely editorial - as \citet{Robinson1996} puts it, as `an able midwife who helped his wife bring her monster to life'? Or was he, as \citet{Rieger1974} suggests, as much as a `minor collaborator'? Or are the fringe theories correct in claiming that Percy Shelley was the real author, and \emph{Frankenstein} his brainchild, rather than his wife's?

This is precisely the question which we seek to answer in this investigation, although we will use a different approach to the above scholars. Rather than concerning ourselves with literary or handwriting analysis, we will instead apply modern techniques of stylometry. To our knowledge, the only previous applications of stylometry to Shelley's work are \citet{Rybicki2016}, a study on authorial gender signals, which analyses whether \textit{Frankenstein} is stylistically similar to the work of other female novelists; and \citet{OSullivan2021}, where selected works of Mary Shelley are found to be similar in style to works from her parents, William Godwin and Mary Wollstonecraft. However, none of these works are explcitly concerned with the authorship of \textit{Frankenstein}, nor do they compare it to works written by Percy Shelley.

\section{Basic Elements of Stylometry}

Stylometry involves using statistical methods to study linguistic style, often to determine the authorship of a particular text of interest. For this purpose, `style' refers to elements such as word choice, sentence structure, and use of punctuation taken together to provide a unique profile for an author. General reviews of the subject and its methods are given in \citet{Juola2006}, \citet{Koppel2009}, \citet{Savoy2020}, and \citet{Stamatatos2009}.

Two widely studied tasks in stylometry are \textbf{Authorship Attribution} which seeks to determine the author of a given text from a (small) pool of potential authors, and \textbf{Authorship Verification}, where we are given a text and a single author and wish to determine (i.e. verify) whether the text was written by said author.

A seminal example of authorship attribution comes from \citet{MostellerWallace1963}, which uses frequency counts of 70 function words to determine the authorship of twelve disputed papers from \emph{The Federalist Papers} between James Madison and Alexander Hamilton. Other examples include identifying J.K. Rowling as the author of \emph{The Cuckoo's Calling}, published under the pseudonym Robert Galbraith \citep{Juola2013,Juola2015}; an analysis of Edgar Allan Poe's disputed writings \citep{Schoberlein2016}; determining whether the \emph{Book of Mormon} was written by a single author or a collaboration \citep{Holmes1992}; determining whether the \emph{Iliad} and the \emph{Odyssey} have the same author \citep{Marindale1996}; verifying the authorship of a poem attributed to Shakespeare \citep{Thisted1987}; other analyses surrounding Shakespeare's authorship \citep{Craig2009,Leigh2019}; and a recent investigation assessing the claim of certain commentators that \emph{Wuthering Heights} was not written by Emily Bront{\"e}, rather her brother Branwell \citep{McCarthy2020}.

\subsection{Function Words}\label{sec:functionwords}

Since the analysis of  \citet{MostellerWallace1963}, function words have played a key role in authorship attribution and verification. These are generally thought taken to be the building blocks of language, and are largely context-free, in the sense that they have essentially no meaning on their own. Function words include articles, prepositions, auxiliary verbs, conjunctions, and pronouns - e.g. `and', `the', `of', and so on. An overview of the suitability and usefulness of function words in stylometry is given in \citet{Argamon2005}, \citet{Garcia2006}, and \citet{Kestemont2014}. 

While certain words should always be considered in a list of function words, such as `the', `a', `and', `it', and so on, there is no single authoritative list of function words that are always used in stylometry. For example, in their analysis of the Federalist Papers, \citet{MostellerWallace1963} choose a particular list of 70 function words. Alternatively, it is common to instead take these words to be the most frequently occuring words in a literary corpus, such as the 100 or 200 most frequent words (MFWs). Once we have a list of function words (or frequent words), variations in their frequencies can be analysed to make comparisons of style between various authors and texts and perform authorship identification.

\subsection{n-grams}

An alternative method for determining the authorship of texts is to focus on character n-grams rather than function words \citep{Keselj2003,Kjell1994}. A character n-gram is a sequence of $n$ consecutive characters. For example, in the sentence `The quick brown fox jumps over the lazy dog', the 3-grams of characters are `the', `he ', `e q', ` qu', and so on. In stylometry, to avoid the choice of character n-gram containing too much context-specific information, a small $n$ is suitable, e.g. 2 or 3 \citep{Stamatatos2009}. However, the actual process of modelling and analysis does not depend on the value of $n$. Indeed, the value of $n$ need not even be fixed \citep{Houvardas2006}. Then, just as for frequent words, variations in the frequencies of n-grams may be analysed to determine authorship.

One of the benefits of n-grams over function words is that, by Zipf's Law \citep{Piantadosi2014,Zipf1936,Zipf1949}, the number of unique words used by an author will be fairly small, so it would be difficult to use as many as 1000 frequent words without encoding large amounts of context-specific information. Since there are significantly more common n-grams since they can span word boundaries, the number of n-grams used can scale up to much higher numbers than is possible with individual words.

In \citet{Grieve2007}, n-grams (with various values of $n$) are found to be some of the most accurate techniques of those tested in author attribution, with 2- and 3-grams being most accurate, and it was also found in \citet{Stamatatos2013} that character n-grams were more reliable than those based on frequent words when comparing works from different genres and/or topics, implying that the n-grams are more independent of context than word-tokens.

\section{Description of Corpus}\label{sec:corpus}

Our goal is to perform authorship attribution on \textit{Frankenstein} by comparing its function word/n-gram frequencies to those of several candidate authors. The full corpus that we use for this is given in \Cref{sec:corpusapp} and contains works by authors such as Percy Shelley and William Godwin (Mary Shelley's father). We now discuss this corpus in more detail, to justify the inclusions which we have made.

In order to construct a represenative authorial signature for Mary Shelley, we include the novels which she wrote during her life: \emph{Valperga}, \emph{The Last Man}, \emph{The Fortunes of Perkin Warbeck}, \emph{Lodore}, \emph{Falkner: A Novel}, and \emph{Mathilda}. These works are particularly important, since if we want to determine whether Mary Shelley wrote \emph{Frankenstein}, we must compare its style to the other books which we know she wrote. If \emph{Frankenstein} is stylistically similar to these undisputed works, we have found evidence that they share an author, i.e. that Mary Shelley wrote \emph{Frankenstein} too.

For our main analysis, we will use the 1818 first edition of \emph{Frankenstein}, as substantive changes were made to the second and third editions in 1823 and 1831 (see \citet{Changes1823} and \citet[Appendix B]{Shelley2019}).\endnote{It is also possible to view the 1818 and 1831 editions side by side here: \url{http://knarf.english.upenn.edu/Text/text.html} (accessed 8 January 2022).} In particular, many of the changes were made by Mary Shelley's father, William Godwin, therefore if we wish to work with the novel in its most unaltered form, it is the 1818 edition which we prefer. 

To use \emph{Frankenstein} as part of our corpus, we make a few initial modifications to the plain text edition which we took from Project Gutenberg. First, we removed all the pre-amble added by Project Gutenberg, including for example their licensing, and we also remove text related to title pages and contents. Furthermore, since the text contains some poems from other authors - such as Percy Shelley's \emph{Mutability} and Wordsworth's \emph{Tintern Abbey} - we remove these too, even if they do not comprise more than a negligible proportion of the overall word count.

For the Percy Shelley texts in our corpus, we include his two early novels, \emph{Zastrozzi}, and \emph{St. Irvyne}. A set of his prose essays was considered for inclusion,\endnote{The particular collection being \emph{A Defence of Poetry and Other Essays}, available at \url{https://www.gutenberg.org/ebooks/5428} (accessed 8 January 2022).} however, these were found via Clustering and Principal Component Analysis to be stylistically very different from the novels, to the extent that considering them as part of his authorial profile led to a loss of accuracy when testing attribution methods. Hence the essays were omitted. This is consistent with the general belief in stylometry that prose and non-prose writings can have quite different authorial signatures even when produced by the same author.

The majority of Percy Shelley's written work is poetry. Unfortunately, this might be unsuitable in characterising authorial style in prose fiction, as noted for example in \citet[note 4]{McCarthy2020}. Therefore, while we have included Percy Shelley's poetry in the corpus (see Appendix for details on the precise editions), it is categorised differently, so that we form two distinct authorial profiles for Percy Shelley: one for poetry and one for prose works.\endnote{In fact, in preliminary analysis where poems were included, these were found via cluster analysis and principal component analysis to appear significantly different from the novels and essays, giving us evidence for this choice.}

Finally for Mary and Percy Shelley, we included the collaborative piece \emph{History of a Six Weeks' Tour}  but considered it as a separate authorial profile independent of those for Mary and Percy as individuals.

For comparison, we include in the corpus some works of related authors: Mary Shelley's parents, William Godwin and Mary Wollstonecraft; the `Godwinian' Charles Brockden Brown; Thomas Love Peacock, who was a close friend of Percy Shelley; gothic author Bram Stoker; Sir Walter Scott, an important figure in the literary world when the Shelleys were writing; and John Polidori, associated with the Romanic movement and author of \emph{The Vampyre}, produced during the same ghost story writing contest from which \emph{Frankenstein} is said to have been conceived. It is also possible to see from Mary's journals\endnote{\url{http://knarf.english.upenn.edu/MShelley/bydates.html} (accessed 8 January 2022).} that she had read some works of certain authors included here - particularly Brown, Godwin, and Scott.

Other than Godwin's \emph{Fleetwood}, all the texts in our corpus are obtained from Project Gutenberg or Project Gutenberg Australia. Some general changes were made for clean-up: Project Gutenberg licenses and other pre-amble were removed, apostrophes are standardised, the underscores used to indicate italic text are removed, as are epigraphs, quotations of other works, introductions, prefaces (including those written by the author), and textual notes.

The R programming language version 4.0.2 was used for our the analysis \citep{RLanguage}, and the \texttt{stylo} package was used for access to certain stylometry and machine learning methods \citep{StyloPackage}. We also implemented some other algorithms - in particular, Delta -  manually. For the analysis with \texttt{stylo}, we used both individual tokens and character 3-grams (the latter being used successfully in \citet{Plakias2008}, \citet{Sapkota2015}, and \citet{Stamatatos2013}), and removed pronouns from the corpus, as these can be too particular to a genre or topic \citep{Pennebaker2011}.

\subsection{Clustering}\label{sec:cluster}

Before carrying out a more formal authorship analysis, it is helpful to visualize the relationship between the texts in this corpus. For this purpose, we use hierarchical clustering  \citep{Eder2017} which represents the corpus as a dendogram by repeatedly clustering together texts which are most similar to each other. The process works as follows: we measure the `distance' between each pair of texts in the corpus, defined as the squared distance between the (normalised)  counts of most frequent words (MFWs) or most frequent n-grams. Then, the two texts with the smallest `distance' are joined into a single cluster node. We then estimate the distance from this new node to all other nodes, and repeat the process, until all texts have been linked. The end result is a single cluster (dendrogram), which is shown in \Cref{fig:cluster3gram,fig:clustertoken} for clustering based on the MFWs and most frequent 3-grams respectively. The black lines link texts which are `close' together under the above distance metric.

We see first that the cluster analysis puts the works of each author together as `sub-clusters' (with only one or two errors), suggesting that it is correctly identifying authorial signatures. It can also be seen that \emph{Frankenstein} is clustered within the cluster containing the other works of Mary Shelley, giving preliminary evidence that she is the true author.

{%
	\centering
	\includegraphics[width=\linewidth]{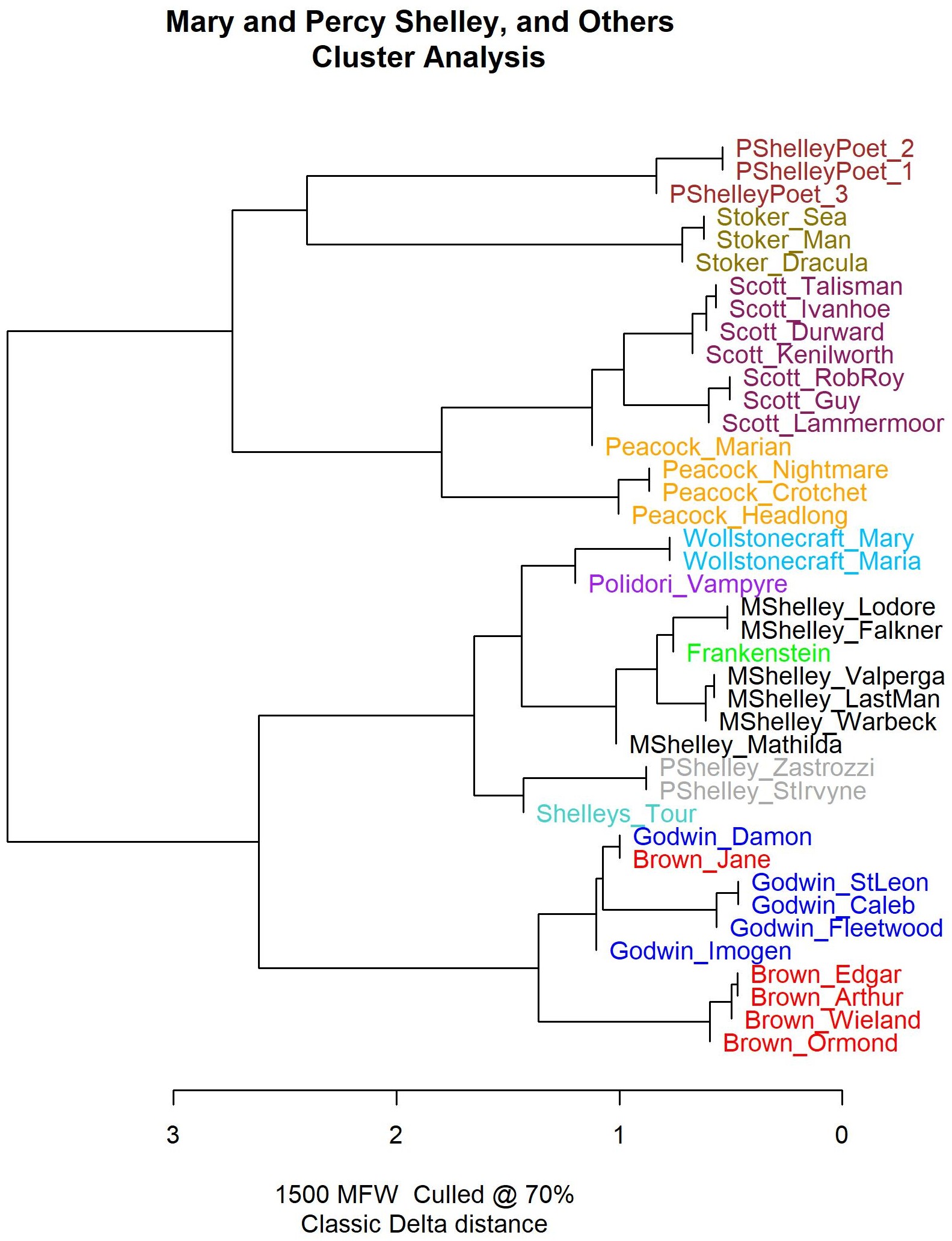}%
	\captionof{figure}{Hierarchical clustering of the corpus  with 70\% culling and using 1500 most frequent 3-grams.}%
	\label{fig:cluster3gram}%
}%
\vspace{1em}
{%
	\centering
	\includegraphics[width=\linewidth]{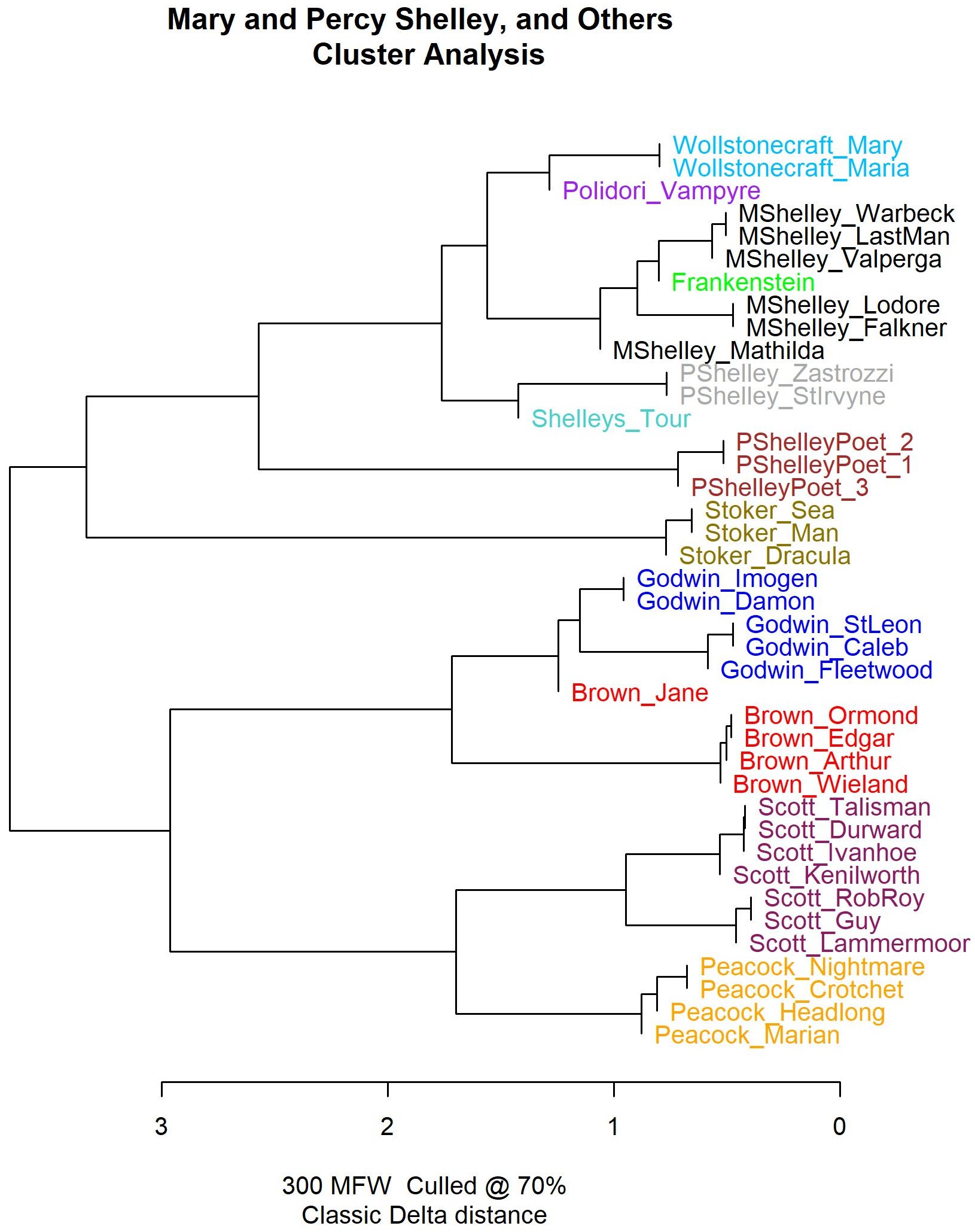}%
	\captionof{figure}{Hierarchical clustering of the corpus with 70\% culling and 300 MFWs.}%
	\label{fig:clustertoken}%
}%

Analysing the exact distances from the distance table produced, we find that on an individual text level, \emph{Frankenstein} is actually closest to \emph{Valperga} - Mary Shelley's novel which immediately followed \emph{Frankenstein} in terms of publication date (\emph{Mathilda} being composed but not published before \emph{Valperga}). This potentially links to the idea of stylistic drift \citep{Ross2020}, where works by a given author that are written closer together in time may also be closer in terms of style.

In general --  and more so in the case of MFWs in \Cref{fig:clustertoken} -- all the works of a given author are properly clustered together. However there are a few exceptions: Charles Brockden Brown's \emph{Jane Talbot} is associated with Godwin's works more than Brown's other novels, and Peacock's \emph{Maid Marian} often occupies the same branch as Sir Walter Scott's works. 

\subsection{Principal Component Analysis}\label{sec:PCAFrank}

We next consider another approach to visualising the corpus, as a prelude to a more formal study. Principal Component Analysis (PCA) is a dimensionality-reduction method, designed to transform a large number of potentially correlated variables into a smaller set of uncorrelated variables, the \emph{principal components} \citep{Binongo1999,Jolliffe2002}. With this process, we hope that much of the variation in the original data can be accounted for in these principal components (PCs). It is one approach to performing Multidimensional Scaling \citep{Borg2005}, i.e. visualising a high dimensional object such as a literary corpus in 2 dimensions, by projecting it down onto its first two principal components.

In the case of stylometry, the high-dimensional variables  are the frequencies of the MFWs or 3-grams, recorded for each text in the corpus. We can use it to visualize in two dimensions some of the stylistic differences encoded in the varying frequencies of the hundreds of features considered.

To determine if PCA allows us to see significant differences between the first edition and the 1831 third edition of \textit{Frankenstein}, we include both in this part of the analysis. Also for the sake of readability (in the visualisation), only a subset of works from the above corpus are considered which represents the most likely candidate authors - those of William Godwin, Mary Shelley, Percy Shelley (both prose and poetry), and the collaborative work between the Shelleys.

The results of the Principal Component Analysis can be seen in \Cref{fig:PCA3gram,fig:PCAtoken}. This plots the various novels in the corpus in 2 dimensions, which is a projection onto the first two principal components. The distance between various texts in this plot is based on the distance between their most frequently occuring words/3-grams, with more similar texts being shown as closer together. Here, we can see a pattern similar to the results of the clustering - the works of each author are fairly closer together geometrically.

{%
	\centering
	\includegraphics[width=\linewidth]{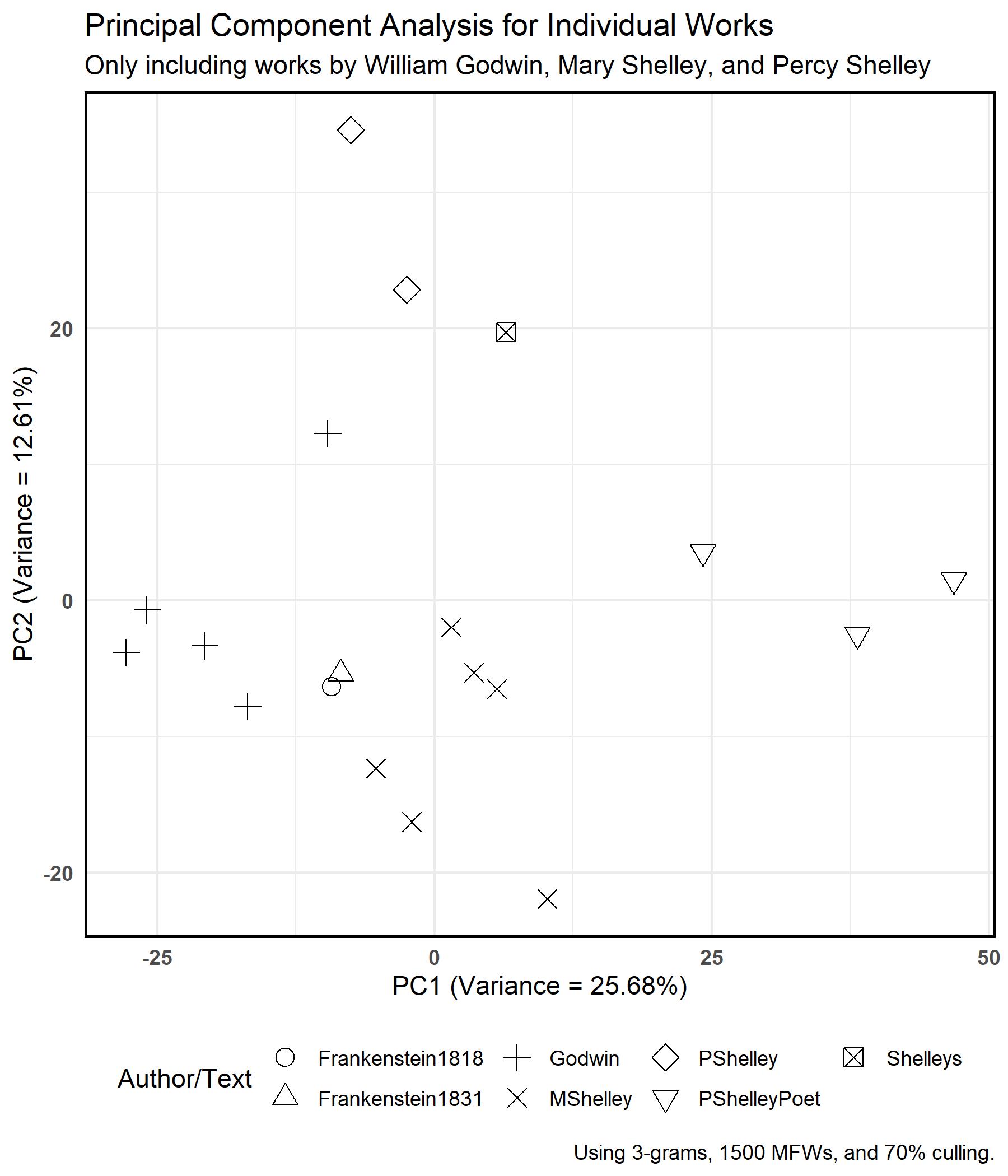}%
	\captionof{figure}{Principal Component Analysis with \emph{Frankenstein} and other works of Mary Shelley, Percy Shelley, and, William Godwin, using the 1500 most frequent 3-grams with $70\%$ culling}%
	\label{fig:PCA3gram}
}%
\vspace{1em}
{%
	\centering
	\includegraphics[width=\linewidth]{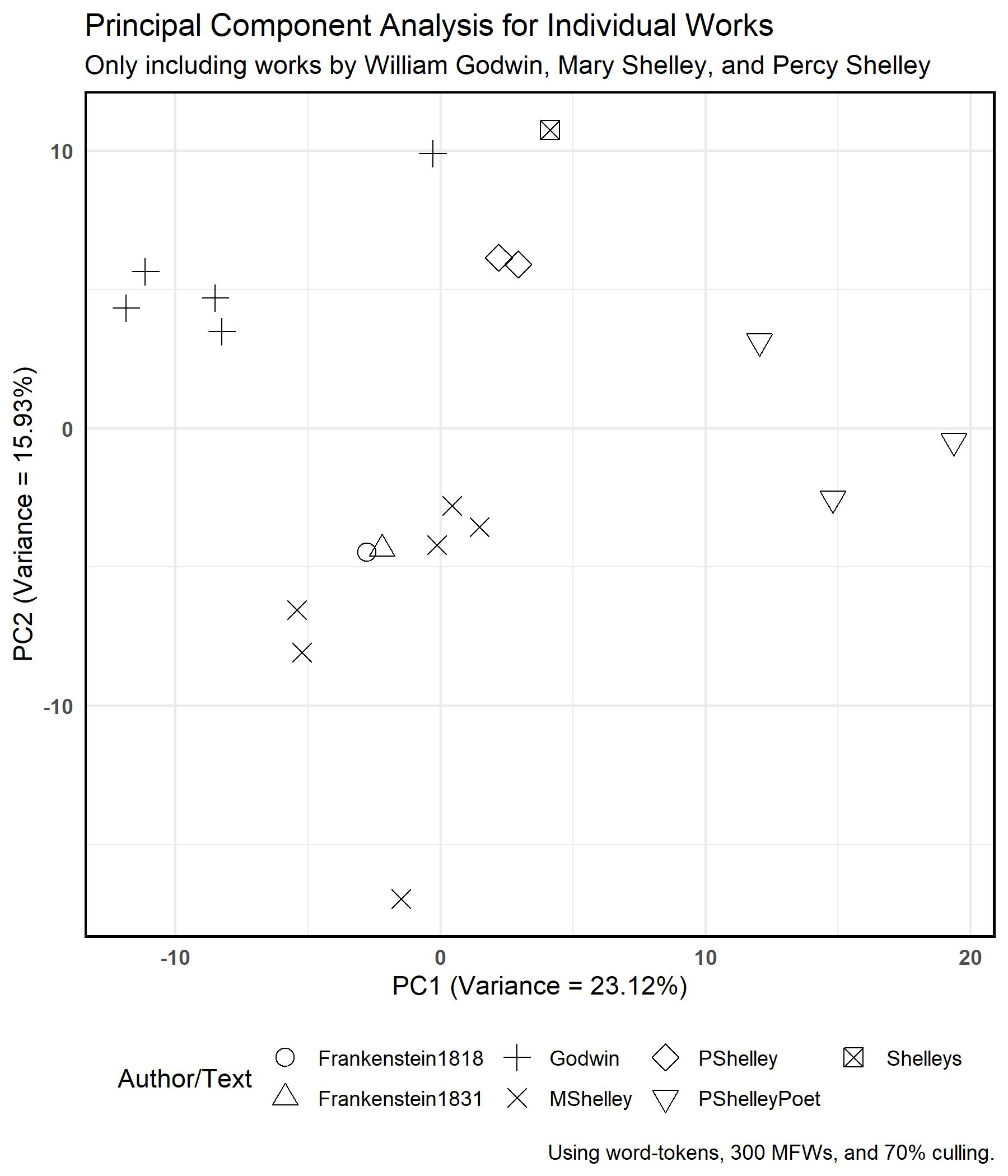}%
	\captionof{figure}{Principal Component Analysis with \emph{Frankenstein} and other works of Mary Shelley, Percy Shelley, and, William Godwin, using the 300 MFWs with 70\% culling.}%
	\label{fig:PCAtoken}
}%
\vspace{1em}

When using MFWs \emph{Frankenstein} is clearly closer to Mary Shelley's other works than to Percy Shelley's or anyone else's. There is more ambiguity when 3-grams are considered, and while it is clear that \emph{Frankenstein} is not at all close to Percy Shelley's work, it is not clear whether it is closer to Mary Shelley's novels, or to William Godwin's. Therefore, while this may provide strong evidence against Percy Shelley's authorship, the evidence in favour of Mary's authorship is not clear-cut. Furthermore, there appears to be little detectable difference between the two editions of \textit{Frankenstein}, so even if there are notable differences between the texts, these differences are in theme and philosophy rather than quantitative style.

A potential concern is that, since \emph{Frankenstein} is written in epistolary form with three distinct narrators - Captain Walton, Victor Frankenstein, and The Creature - it is possible that each narrator has a distinct writing style and that the style associated with \emph{Frankenstein} is the average of three individual stylistic profiles. To check this, PCA was performed in the same way as before, with the parts of each narrator being treated as individual texts. We find that the three data points cluster very close to the novel as a whole, closer than any other text in the corpus in fact, therefore this epistolary form appears to not be a concern.

\section{Methodology}

Having now completed the preliminary visualisation of the corpus, we move on to performing a more formal authorship attribution study. Our goal is to determine which of the authors in our corpus is the most likely author of \textit{Frankenstein}. For this purpose, we will use two standard methods of authorship attribution which are common in the stylometry literature -- Burrows' Delta, and Support Vector Machines.

\subsection{Burrows' Delta}\label{sec:methodology}

The `Delta' method introduced by \citet{Burrows2002}, is essentially a type of K-nearest neighbours classifier which measures the `distance' between the text which we wish to analyse and each author from a set of candidate authors, based on their use of MFWs (or n-grams). More formally, suppose our features consist of a set of $M$ most frequent words or ngrams. For a given candidate author $a$, let $A_a$ be a length $M$ vector which constitutes the profile of that author, which is formed by computing the (normalised) proportion of frequent words used by that author in the corpus texts, (so that the first element of $A_a$ is the proportion of times that the most frequent word appeared in the set of corpus texts written by author $a$, and so on). Next let $C$ denote the text to be classified (e.g. Frankenstein) where again $C$ is a length $M$ vector of frequent word proportions. Finally let $d()$ be a distance function such as Manhattan, Euclidean or Cosine distance. Then, delta is given by:

\begin{equation*}
	\Delta(A_a C) = d(A_a, C)
\end{equation*}

The author which produces the minimum distance score is then identified as the most likely author of the text. Certain modifications of this algorithm are introduced in \citet{Hoover2004a}, \citet{Hoover2004b}, and \citet{Argamon2008}. The main modification used in our investigation are removal of pronouns from the set of frequent words \citep{Hoover2004a} and \textbf{culling}, where words that don't appear in at least a given proportion of texts (e.g. 70\%) are removed. This should remove some words that are too context-specific, or words that are very specific to a single text.

\subsection{Support Vector Machines}

Support vector machines (SVM) are an approach to supervised that is commonly used in stylometry. Given a $M$ dimensional feature set, it aims to find a hyperplane (border) in $M$-dimensional space which distinctly classifies data points from two distinct classes  \citep{Diederich2003,Joachims1998}. Given that multiple possible hyperplane are likely to exist, the standard SVM approach chooses the one which maximises a type of distance (the `margin') between the classes\citet{Cortes1995}. If it is not possible to perfectly separate the classes, a loss function may be used to penalize data points that lie on the wrong side of the hyperplane, and non-linear kernels can be used to project the data into a space where such a separation is possible. 

In applications to stylometry, the data points used in the SVM algorithm will be the corpus texts and the unknown text to be classified, all represented in feature-space as vectors of function word proportions, and the classes represent the candidate authors. Since there will typically be more than two candidate authors, there are two commonly used approaches to convert SVMs into a multi-class classifier., The first is one-versus-all, which pits one class/author against all others; the second is one-versus-one, where binary comparisons are made between every possible pair of classes. In both cases, the multiple binary classification results are aggregated in a kind of `democratic process' to return a single class result. It is the one-versus-one method that is implemented in the \texttt{stylo} R package for authorship attribution, which we use in our analysis.

\section{Results}

\subsection{Delta}\label{sec:delta}

We apply Delta using function word counts of 100, 200, 500, and 1000, both with and without culling at the 70\% level.  For robustness, we consider both the  Manhattan and cosine distance functions. Note that cosine distance is in this case defined as one minus the cosine similarity \citep{cosinesim}, where for two vectors $\bm{x}$ and $\bm{y}$:
\[
\begin{split}
	\text{similarity} &= \cos\theta = \frac{\bm{x}\cdot \bm{y}}{\left\lVert\bm{x}\right\rVert \left\lVert\bm{y}\right\rVert} \\
	&= \frac{\sum_{i=1}^n x_i y_i}{\sqrt{\sum_{i=1}^n x_i^2}\sqrt{\sum_{i=1}^n y_i^2}}.
\end{split}
\]

Note that Delta uses a profile-baeed approach  where all works written by each author are combined into a single data point. The only exception to this is for the works of Percy Shelley, as mentioned in \Cref{sec:corpus}, where we split his works into two profiles: one for poetry and one for prose.

We applied Delta to the corpus using a variety of different feature sets, i.e. different numbers of MFWs and 3g-rams. In all cases, the algorithm attributes \emph{Frankenstein} to Mary Shelley, often by a considerable margin when one compares the raw distance scores, as shown in \Cref{table:deltaclassic,table:deltacosine}. For example, in the case of 200 MFWs (listed in Appendix B), 70\% culling and Manhattan distances, the Delta score for comparison between Mary Shelley's other works is 0.642, compared to 0.840 for the second-nearest author (William Godwin) and 0.938 for Percy Shelley (prose profile). Therefore, in the language of Burrows' original paper, \emph{Frankenstein} is a lot `less unlike' Mary Shelley's other novels than any other authorial profiles we have created \citep{Burrows2002}.

As can be seen, this attribution is independent of which distance metric is used and the number of frequent words employed, which demonstrates robustness (i.e. the results are not sensitive to small variations in how we set up the analysis). In other experiments, we also found the attribution to Mary Shelley to be unchanged with the presence or absence of culling. Within the tables for the two distance metrics, we can see that the general pattern does not significantly vary as number of MFWs varies. For example, with Manhattan distances, Mary Shelley is always ranked first; Godwin, Brown, and Wollstonecraft rank near the top; Percy Shelley, Scott, and Peacock are central; and Polidori, Shelleys, and Stoker are at the bottom of the rankings.

\subsection{SVMs}

We now perform the same authorship attribution using SVMs rather than Delta. For this purpose we use the SVM implementation from the \texttt{stylo} R package. Classification of \emph{Frankenstein} is performed in several ways in order to verify robustness. This is done with two different training sets - one consisting of the entire corpus, and one which reduces the classification problem to a pure binary classification by only considering the works of Mary Shelley and Percy Shelley (Prose). Recall that \texttt{stylo} uses a one-versus-one approach for the multiclass problem. 

To demonstrate that the attribution of \emph{Frankenstein} is not sensitive to the parameters chosen for the model, we perform the analysis on both MFWs and 3-grams, with and without culling at the 70\% level, and with a variety of MFWs. For 3-grams, we used $200, 300, \ldots, 2000$ features, and $50, 100, \ldots, 500$ for MFWs This gives a total of 116 variations.

In all cases, \emph{Frankenstein} is attributed to Mary Shelley, both over Percy Shelley and over any of the other authors in the corpus. Combined with the above results from Delta, this provides extremely strong evidence that Mary Shelley is indeed the author of  \emph{Frankenstein}

\section{Further Robustness Checks}

The above analysis showed that both Delta and SVMs attributed Frankenstein to Mary Shelley. However for this result to be convincing,  we should verify that Delta/SVMs are actually able to correctly identify the author of 19th century literary texts. Although there is a large amount of existing evidence showing the general effectiveness of these algorithms in stylometry, we will now demonstrate for completeness that these are also effective on the corpus we have constructed, which increases the strength of our results.

\subsection{Cross Validation}\label{sec:deltacv}

We use Leave One Out Cross Validation \citep{CrossVal} to check that Delta is able to correctly identify the author of literary texts. For this purpose, we attempt to classify each text in the corpus one by one, by removing it from the corpus, and treating it is as it were the unknown text, with the remaining corpus texts used to form the author profiles. In other words, our goal is to check that our methods can correctly identify the author of each text, and hence demonstrate their reliability. Note two of the authors (John Polidori, and Percy Shelley's poetry works) only had a single work in the corpus so we omitted these from the cross-validation exercise since the resulting training set would not have had any works left by these authors  from which to learn their profile.

After doing this, we found that Delta correctly identified the author of every text in the corpus, with the exception of Thomas Love Peacock's \emph{Maid Marian} which was incorrectly attributed to Sir Walter Scott. This is similar to the result of clustering in  Section \Cref{fig:cluster3gram}, where this novel was clustered closer to Scott's novels than Peacock's other works. Despite thie misclassification, the accuracy of Delta on the corpus is $96\%$ which shows that it is extremely reliable for attributing 19th century literary texts, further increasing the believability of the attribution of Frankenstein to Mary Shelley.

Next we perform a similar analysis to verify the accuracy of SVMs. We found that in this case, there were 4 misclassifications in the corpus, giving an accuracy of $79\%$. The incorrect classifications were Thomas Love Peacock's \emph{Maid Marian} being attributed to Sir Walter Scott; Percy Shelley's \emph{St. Irvyne} to Mary Shelley, and both of Mary Wollstonecraft's novels to Mary Shelley. In addition, \emph{Zastrozzi} and one volume Percy Shelley's poetry are attributed to his wife. This is potentially due to class imbalance - as seen in \Cref{fig:cluster3gram,fig:clustertoken}, Mary and Percy Shelley and Mary Wollstonecraft had fairly similar writing styles compared to other authors in the corpus, and with Mary Shelley having six novels in her profile, this could skew attributions towards her. The mistakes surrounding Percy Shelley's works are particularly concerning in the context of the \emph{Frankenstein} authorship question, so the SVM trained on the corpus should not be given as much weight in concluding the novel's true author. However Delta's near-perfect accuracy shows that it is a reliable method, and fortunately both Delta and SVMs both agree that Mary Shelley is the most likely author of Frankenstein.

\subsection{Impostors Method}

The above analysis focused on authorship \textbf{attribution}, i.e. identifying the most likely author from a given set of candidate authors. However the problem can also be phrased in terms of authorship \textbf{verification} where the question is a binary yes/no decision as to whether a particular author is likely to have written a text. We will now study \emph{Frankenstein} as a verification problem, using the recently introduced `Imposters' algorithm. 

The Imposters method was introduced by \citet{Koppel2014} and has since been applied to several problems in stylometry, for exmaple in \citet{Kestemont2016}. We now provided a brief summary of the method, and the interested reader can consult \citet{Koppel2014} for a full description. Suppose we have an unknown text (e.g. \emph{Frankenstein}) represented as a vector $\mathbf{x}$ of features and we wish to determine if it was written by a particular candidate author. Let $T = \{t_1, \ldots, t_n\}$ denote a set of texts which are known to be written by this author. Next, assume we have a set of distractor documents written by other authors (i.e. the rest of the corpus) known as `impostors', $I = \{i_1, \ldots, i_k\}$. During each iteration of the algorithm (typically 100 in total), the method selects a random subset of the impostors, $I'$, and determines whether $\mathbf{x}$ is closer to an item in $T$ than in $I'$, considering a proportion (e.g. 0.5) of the total features available. The value returned by the method is a value between 0 and 1 representing the proportion of iterations for which the analysed text was closer to a work of the suspected author than one of the impostors (i.e. closer to an element of $T$ than one of $I'$). If the returned score is larger than a threshold $\sigma$, then we have verified that the text was written by this particular author. Otherwise, verification has failed.

Similar to the discussions mentioned in the section of this report on the Delta framework, the choice of distance/similarity metric may significantly affect the performance of the algorithm. Popular choices are the Manhattan distance, the cosine distance, and the minmax distance, as used by \citet{Koppel2014}, where for two vectors $\mathbf{x}$ and $\mathbf{y}$:
\begin{equation*}
	\text{minmax}(\mathbf{x},\mathbf{y}) = \frac{\sum_{i=1}^n \text{min}(x_i, y_i)}{\sum_{i=1}^n \text{max}(x_i, y_i)},
\end{equation*}
where $x$ and $y$ are vectors of $n$ features. Distances between z-scores instead of relative frequencies may also be considered (`Delta' distances).

We apply the impostors method to \emph{Frankenstein} as a test of robustness, using the \texttt{imposters()} function from the \texttt{stylo} R package, considering both word-tokens and 3-grams. For simplicity, we begin with using default values; that is, 100 iterations, selection of 50\% of texts as impostors and 50\% of features for comparisons, and the classic Delta distance.

The \texttt{R} output from running the impostors method is given in \Cref{fig:impostors}. Note that the numerical values are not fixed due to the stochastic nature of the algorithm.

\begin{figure*}
	\centering
	
	\hspace{0.18\textwidth}
	\begin{subfigure}[b]{0.24\textwidth}
		\centering
		\begin{verbatim}
			Brown 	 0
			Godwin 	 0.07
			MShelley 	 1
			Peacock 	 0
			Polidori 	 0
			PShelley 	 0
			PShelleyPoet 	 0
			Scott 	 0
			Shelleys 	 0
			Stoker 	 0
			Wollstonecraft 	 0
		\end{verbatim}
		\caption{3-grams}
	\end{subfigure}
	\hfill
	\begin{subfigure}[b]{0.24\textwidth}
		\centering
		\begin{verbatim}
			Brown 	 0
			Godwin 	 0.05
			MShelley 	 1
			Peacock 	 0
			Polidori 	 0
			PShelley 	 0
			PShelleyPoet 	 0
			Scott 	 0
			Shelleys 	 0
			Stoker 	 0
			Wollstonecraft 	 0
		\end{verbatim}
		\caption{word-tokens}
	\end{subfigure}
	\hspace{0.18\textwidth}
	
	\caption{Results from applying the Impostors algorithm to \emph{Frankenstein}, using the classic Delta distance.}
	\label{fig:impostors}
\end{figure*}

In either case, the score of 1 for Mary Shelley (\texttt{MShelley}) is unambiguous - the algorithm does not find any reason to doubt that Mary Shelley wrote \emph{Frankenstein}. In technical terms, during every iteration, \emph{Frankenstein} is found to be closer to one of Mary Shelley's novels than to any of the impostor works. While being different from zero, the scores for William Godwin of 0.07 and 0.05 still indicate that we should reject Godwin as the novel's author. With regards to the original hypothesis of \emph{Frankenstein} being Percy Shelley's composition, we find no evidence in favour of this claim here.

We can also use \texttt{imposters.optimize()} from \texttt{stylo} to tune the thresholds for a `yes' and `no' decision. We find the lower \texttt{p1} values to be 0.38 and 0.20 for 3-grams and tokens, respectively, indicating that a value lower than that may be taken as a rejection of that person as the author. Similarly, the \texttt{p2} values that give the cut-off for a `yes' decision are found to be 0.61 and 0.78 - these are the values of $\sigma$ referenced earlier. Note that between the \texttt{p1} and \texttt{p2} values is the `grey area' for the algorithm - we cannot reliably draw a conclusion as to predicted authorship. Fortunately, none of the returned scores fall within this range in our case.

To check our results for robustness, we also ran the impostors algorithm using various different distance metrics and choices of MFWs and 3-grams. Attribution to Mary Shelley, combined with a notable lack of evidence in favour of Percy Shelley's authorship, occurs in every case. As before, Godwin is identified as a more likely candidate than Percy Shelley, similar to what we found when looking at Delta distances in \Cref{sec:delta}.

Note that the impostors method is likely to be sensitive to some authors having more works in the corpus than others since the algorithm makes instance-based comparisons. Therefore, it is possible that, for example, Scott would be put at an advantage with six works, over Polidori and Wollstonecraft with only one or two works. This is the usual problem of class imbalance which can affect the problem of some supervised learning methods.

To gain additional confidence that the attribution to Mary Shelley is not overly influenced by class imbalance, we apply the impostors method to all other works from Mary and Percy Shelley for which authorship is undisputed, of which there are eleven in total. For example, we will confirm that \emph{Zastrozzi} is attributed to Percy Shelley, even if he has fewer works making up the authorial profile when instance-based comparisons are considered. In doing this, the work in question is removed from the corpus of works for comparison, and we work only with the data on the frequency of MFWs. In other words, we are performing leave on out cross-validation.

The result is as we would hope - for all works, we achieve a score from the algorithm of 1 for the expected author. For Percy Shelley's works, there are non-trivial scores for Mary Shelley's potential authorship, but in the application to Mary Shelley's novels, all receive scores of 0 as far as Percy Shelley's authorship goes. Therefore, in summary, when dealing with the novels, the results from the impostors algorithm appear consistent and reliable.

\section{Limitations}

While we believe our above analysis is fairly conclusively, we should mention potential limitations of our approach. One objection is that we have assumed that all Mary Shelley's post-\emph{Frankenstein} novels were actually written by her. However, it is theoretically possible that Percy Shelley may have had a major role in all of her novels, in which case our analysis would not be valid. However this seems highly unlikely, and to our knowledge has never been suggested by any literary scholar. Indeed, as discussed earlier, it was the apparent stylistic difference between \emph{Frankenstein} and Mary's second novel which initial fueled speculation that Percy may have contributed to \emph{Frankenstein}. Additionally,  Percy Shelley died in July 1822, it is only possible for him to have contributed to \emph{Mathilda} and potentially \emph{Valperga} also.

A second potential limitation is the relatively small number of Percy Shelley prose works included in the corpus. This mostly stems from the fact that he was not primarily a novelist, and while his works of poetry and prose essays were included, both are seen to be dissimilar to the novels, leading to the poetry only being included as forming a second authorial profile for Shelley. However our cross-validation study has confirmed the general finding in the stlyometric community that Delta is able to accurately identify authorship even given only a relatively small number of available works, and so we do not believe that this is a major issue.

\section{Conclusion}

In conclusion, the above analysis points strongly in favour of Mary Shelley's authorship for \emph{Frankenstein}. This finding is robust against several different types of stylometric method including classic authroship attribution algorithms such as Delta and SVMs, along with the unsupervised Imposters approach for authorship verification. Our robutness studies showed that these techniques have high accuracy for attributing  the authorship of 19th century literary texts, and can hence be trusted.  While it is known from \citet{Robinson2008} that Percy's contribution to \emph{Frankenstein} was not insignificant, his style is simply not close enough to \emph{Frankenstein} to suggest that he was responsible for the novel as a whole, as some have suggested \citep{deHart2013,JonesUnpublished,Lauritsen2007,Zimmerman1998}.

To the (small) extent that a second person is implicated as a potential author of \emph{Frankenstein}, it is William Godwin who prevails over Percy Shelley, for example in the Principal Component Analysis applied to 3-grams, and in our Imposters study. This closeness of Godwin's style to that adopted by Mary Shelley is not altogether surprising given that he is her father and was responsible for her (somewhat informal) education. This potentially also links to the `parental meddling' suspected in \citet[note 2]{Rybicki2016}. Further research could be done building on this study and \citet{OSullivan2021} to investigate the influence of Godwin on his daughter's style, both in his involvement in her education and from an editorial perspective.

\section*{Acknowledgements}

"We thank Daragh Meehan, who completed an earlier University of Edinburgh honours dissertation on the same supervisor-proposed topic, for allowing that dissertation to be shared during the project.

\theendnotes

\bibliographystyle{dsh}
\bibliography{sources}

\end{multicols}

\afterpage{
	\clearpage
	\thispagestyle{empty}
	\begin{landscape}
		\begin{table}
			\centering
			\begin{tabular}{llc|llc|llc|llc}
				\multicolumn{3}{c|}{100 MFWs} & \multicolumn{3}{|c|}{200 MFWs} & \multicolumn{3}{|c|}{500 MFWs} & \multicolumn{3}{|c}{1000 3-grams} \\
				Rank & Author & Distance & Rank & Author & Dist. & Rank & Author & Dist. & Rank & Author & Dist. \\
				\hline
				1 & Mary Shelley & 0.578              & 1 & M. Shelley & 0.642           & 1 & M. Shelley & 0.688          & 1 & M. Shelley & 0.768 \\
				2 & William Godwin & 0.815            & 2 & Godwin & 0.840               & 2 & Godwin & 0.892              & 2 & Godwin & 0.900 \\
				3 & Charles Brockden Brown & 0.880    & 3 & Brown & 0.870                & 3 & Wollstonecraft & 0.938      & 3 & Wollstonecraft & 0.979 \\
				4 & Walter Scott & 0.899              & 4 & Wollstonecraft & 0.886       & 4 & Brown & 0.943               & 4 & Brown & 0.988 \\
				5 & Mary Wollstonecraft & 0.955       & 5 & P. Shelley (Prose) & 0.938   & 5 & P. Shelley (Prose) & 1.014  & 5 & Scott & 1.087 \\
				6 & Percy Shelly (Prose) & 0.996      & 6 & Scott & 1.010                & 6 & Scott & 1.036               & 6 & P. Shelley (Prose) & 1.095 \\
				7 & Percy Shelley (Poetry) & 1.048    & 7 & P. Shelley (Poetry) & 1.027  & 7 & Peacock & 1.071             & 7 & Peacock & 1.138 \\
				8 & Thomas Love Peacock & 1.085       & 8 & Peacock & 1.111              & 8 & P. Shelley (Poetry) & 1.130 & 8 & P. Shelley (Poetry) & 1.209 \\
				9 & John Polidori & 1.087             & 9 & Polidori & 1.151             & 9 & Polidori & 1.160            & 9 & Shelleys & 1.249 \\
				10 & Bram Stoker & 1.162              & 10 & Shelleys & 1.216            & 10 & Shelleys & 1.211           & 10 & Stoker & 1.250 \\
				11 & Shelleys (Collaboration) & 1.296 & 11 & Stoker & 1.277              & 11 & Stoker & 1.255             & 11 & Polidori & 1.264 \\
			\end{tabular}
			\captionof{table}{Delta Scores for classification of \emph{Frankenstein} with 70\% culling, and Manhattan distance (Classic Delta). Lower numbers indicate that the author is more likely to have written the text.}
			\label{table:deltaclassic}
		\end{table}
		
		\begin{table}
			\centering
			\begin{tabular}{llc|llc|llc|llc}
				\multicolumn{3}{c|}{100 MFWs} & \multicolumn{3}{|c|}{200 MFWs} & \multicolumn{3}{|c|}{500 MFWs} & \multicolumn{3}{|c}{1000 3-grams} \\
				Rank & Author & Distance & Rank & Author & Dist. & Rank & Author & Dist. & Rank & Author & Dist. \\
				\hline
				1 & Mary Shelley & 0.598              & 1 & M. Shelley & 0.633          & 1 & M. Shelley & 0.592          & 1 & M. Shelley & 0.631 \\
				2 & Charles Brockden Brown & 0.819    & 2 & Brown & 0.850               & 2 & Wollstonecraft & 0.896      & 2 & Godwin & 0.846 \\
				3 & William Godwin & 0.887            & 3 & Wollstonecraft & 0.876      & 3 & P. Shelley (Prose) & 0.924  & 3 & Wollstonecraft & 0.846 \\
				4 & Bram Stoker & 0.959               & 4 & Godwin & 0.923              & 4 & Brown & 0.941               & 4 & Brown & 0.925 \\
				5 & Percy Shelley (Prose) & 0.984     & 5 & P. Shelley (Prose) & 0.951  & 5 & Godwin & 0.952              & 5 & P. Shelley (Prose) & 1.930 \\
				6 & Mary Wollstonecraft & 1.005       & 6 & P. Shelley (Poetry) & 0.960 & 6 & Shelleys & 0.969            & 6 & Shelleys & 1.004 \\
				7 & Percy Shelley (Poetry) & 1.010    & 7 & Shelleys & 1.027            & 7 & Polidori & 0.987            & 7 & Polidori & 1.023 \\
				8 & John Polidori & 1.078             & 8 & Polidori & 1.031            & 8 & P. Shelley (Poetry) & 1.010 & 8 & P. Shelley (Poetry) & 1.040 \\
				9 & Walter Scott & 1.136              & 9 & Stoker & 1.101              & 9 & Stoker & 1.168              & 9 & Stoker & 1.190 \\
				10 & Shelleys (Collaboration) & 1.140 & 10 & Scott & 1.221              & 10 & Scott & 1.230              & 10 & Scott & 1.205 \\
				11 & Thomas Love Peacock & 1.213      & 11 & Peacock & 1.248            & 11 & Peacock & 1.238            & 11 & Peacock & 1.251 \\
			\end{tabular}
			\captionof{table}{Delta Scores for classification of \emph{Frankenstein} with 70\% culling, and cosine distance. Lower numbers indicate that the author is more likely to have written the text.} 
			\label{table:deltacosine}
		\end{table}
	\end{landscape}
	\clearpage
}

\pagebreak
\appendix
\section{Composition of Corpus}\label{sec:corpusapp}

\hypersetup{urlcolor=blue}

In the following table, \href{https://www.gutenberg.org/}{PG} = Project Gutenberg; \href{http://gutenberg.net.au/}{PGA} = Project Gutenberg Australia; and \href{https://theanarchistlibrary.org/special/index}{AL} = The Anarchist Library. Note that while \emph{Mathilda} was only published posthumously in 1959, it was written in 1819 \citep{ShelleyMathilda,Fisch1993}, thus the year is recorded as 1819.

\begin{table}[H]
	\centering
	\begin{tabular}{cllcc}
		Number & Author & Title & Year & Source \\
		\hline
		1 & To be Determined & Frankenstein & 1818, 1831 & \href{https://www.gutenberg.org/ebooks/41445}{PG}, \href{https://www.gutenberg.org/ebooks/84}{PG} \\
		2 & Charles Brockden Brown & Arthur Mervyn & 1799 & \href{https://www.gutenberg.org/ebooks/18508}{PG} \\
		3 & Charles Brockden Brown & Edgar Huntly & 1799 & \href{https://www.gutenberg.org/ebooks/8223}{PG} \\
		4 & Charles Brockden Brown & Jane Talbot & 1801 & \href{https://www.gutenberg.org/ebooks/8404}{PG} \\
		5 & Charles Brockden Brown & Ormond & 1799 & \href{http://gutenberg.net.au/ebooks06/0605271.txt}{PGA} \\
		6 & Charles Brockden Brown & Wieland & 1798 & \href{https://www.gutenberg.org/ebooks/792}{PG} \\
		7 & William Godwin & Caleb Williams & 1794 & \href{https://www.gutenberg.org/ebooks/11323}{PG} \\
		8 & William Godwin & Damon and Delia & 1784 & \href{https://www.gutenberg.org/ebooks/10318}{PG} \\
		9 & William Godwin & Fleetwood & 1805 & \href{https://theanarchistlibrary.org/library/william-godwin-fleetwood}{AL} \\
		10 & William Godwin & Imogen & 1784 & \href{https://www.gutenberg.org/ebooks/9152}{PG} \\
		11 & William Godwin & St. Leon & 1799 & \href{https://www.gutenberg.org/ebooks/53707}{PG} \\
		12 & Thomas Love Peacock & Crotchet Castle & 1831 & \href{https://www.gutenberg.org/ebooks/2075}{PG} \\
		13 & Thomas Love Peacock & Headlong Hall & 1815 & \href{https://www.gutenberg.org/ebooks/12803}{PG} \\
		14 & Thomas Love Peacock & Maid Marian & 1822 & \href{https://www.gutenberg.org/ebooks/966}{PG} \\
		15 & Thomas Love Peacock & Nightmare Abbey & 1818 & \href{https://www.gutenberg.org/ebooks/9909}{PG} \\
		16 & John Polidori & The Vampyre & 1819 & \href{https://www.gutenberg.org/ebooks/6087}{PG} \\
		17 & Mary Shelley & Falkner & 1837 & \href{http://gutenberg.net.au/ebooks06/0603201.txt}{PGA} \\
		18 & Mary Shelley & Lodore & 1835 & \href{http://gutenberg.net.au/ebooks06/0606381.txt}{PGA} \\
		19 & Mary Shelley & Mathilda & 1819 & \href{https://www.gutenberg.org/ebooks/15238}{PG} \\
		20 & Mary Shelley & The Fortunes of Perkin Warbeck & 1830 & \href{http://gutenberg.net.au/ebooks06/0606411.txt}{PGA} \\
		21 & Mary Shelley & The Last Man & 1826 & \href{https://www.gutenberg.org/ebooks/18247}{PG} \\
		22 & Mary Shelley & Valperga & 1823 & \href{http://gutenberg.net.au/ebooks06/0606801.txt}{PGA} \\
		23 & Percy Shelley & St. Irvyne & 1811 & \href{http://gutenberg.net.au/ebooks06/0606391.txt}{PGA} \\
		24 & Percy Shelley & Zastrozzi & 1810 & \href{http://gutenberg.net.au/ebooks06/0606461.txt}{PGA} \\
		25-27 & Percy Shelley (Poetry) & Complete Poetical Works (3 Vols) & Various & \href{https://www.gutenberg.org/ebooks/4800}{PG} \\
		28 & Sir Walter Scott & Guy Mannering & 1815 & \href{https://www.gutenberg.org/ebooks/5999}{PG} \\
		29 & Sir Walter Scott & Ivanhoe & 1819 & \href{https://www.gutenberg.org/ebooks/82}{PG} \\
		30 & Sir Walter Scott & Kenilworth & 1821 & \href{https://www.gutenberg.org/ebooks/1606}{PG} \\
		31 & Sir Walter Scott & Quentin Durward & 1823 & \href{https://www.gutenberg.org/ebooks/7853}{PG} \\
		32 & Sir Walter Scott & Rob Roy & 1817 & \href{https://www.gutenberg.org/ebooks/7025}{PG} \\
		33 & Sir Walter Scott & The Talisman & 1825 & \href{https://www.gutenberg.org/ebooks/1377}{PG} \\
		34 & Mary and Percy Shelley & History of a Six Weeks' Tour & 1817 & \href{https://www.gutenberg.org/ebooks/52790}{PG} \\
		35 & Bram Stoker & Dracula & 1897 & \href{https://www.gutenberg.org/ebooks/345}{PG} \\
		36 & Bram Stoker & The Man & 1905 & \href{https://www.gutenberg.org/ebooks/2520}{PG} \\
		37 & Bram Stoker & The Mystery of the Sea & 1902 & \href{https://www.gutenberg.org/ebooks/42455}{PG} \\
		38 & Mary Wollstonecraft & Maria, or The Wrongs of Woman & 1798 & \href{https://www.gutenberg.org/ebooks/134}{PG} \\
		39 & Mary Wollstonecraft & Mary: A Fiction & 1788 & \href{https://www.gutenberg.org/ebooks/16357}{PG} \\
	\end{tabular}
\end{table}

\newpage
\section{200 words}\label{sec:200app}

\begin{figure}[H]
	\begin{displayquote}
		the, i, to, and, a, of, in, that, it, my, is, you, for, was, on, me, but, so, this, with, have, be, we, at, not, all, im, as, like, are, just, its, out, up, about, they, what, or, one, if, do, from, had, get, when, will, there, don't, time, know, now, can, some, then, by, really, no, well, an, your, go, more, were, am, think, would, who, people, good, been, how, has, got, them, going, because, back, day, see, much, our, only, which, want, their, love, even, other, too, after, today, went, over, way, here, last, into, ive, still, say, could, very, things, new, did, life, work, off, something, right, make, than, us, first, that's, said, didn't, little, cant, night, down, never, thing, also, why, again, around, being, feel, ill, home, where, any, should, need, take, two, before, most, while, those, oh, come, these, made, great, though, long, always, better, ever, friends, myself, another, many, since, next, maybe, thought, look, through, fun, few, bad, actually, find, world, week, lot, sure, days, year, someone, man, pretty, years, away, getting, every, tell, may, best, god, came, anything, same, nothing, let, stuff, doing, read, old, everyone, guess, place, put, nice, left, own, told
	\end{displayquote}
	\caption{200 words.}
\end{figure}

\end{document}